\documentstyle[astrop-bib]{mn}
\input{epsf}

\newcommand{\acir}{$\alpha$~Cir}
\newcommand{\gequ}{$\gamma$~Equ}
\newcommand{\hr}{HR~3831}
\newcommand{\halp}{H$\alpha$}
\newcommand{\ms}{\,m\,s$^{-1}$}
\newcommand{\mh}{\,$\mu$Hz}
\newcommand{\rcw}{$R_{cw}$}

\title[Time-series spectroscopy of \hr]
{Time-series spectroscopy of the rapidly oscillating Ap star \hr}

\author[I.~K.~Baldry and T.~R.~Bedding]
{Ivan~K.~Baldry$^{1,2}$ and Timothy~R.~Bedding$^1$\\
$^1$Chatterton Astronomy Department, School of Physics,
University of Sydney, NSW 2006, Australia\\
$^2$Anglo-Australian Observatory, P.O.\,Box 296, Epping, NSW 1710, Australia}

\date{Accepted 2000 April. Received 2000 April; in original form 2000 January}

\pagerange{\pageref{firstpage}--\pageref{lastpage}}

\pubyear{2000}

\begin{document}

\maketitle
\label{firstpage}

\begin{abstract}
We present time-series spectroscopy of the rapidly oscillating Ap star \hr.
This star has a dominant pulsation period of 11.7 minutes and a rotation
period of 2.85\,d.  We have analysed 1400 intermediate-resolution spectra
of the wavelength region 6100--7100\AA\ obtained over one week, using
techniques similar to those we applied to another roAp star, \acir.

We confirm that the \halp\ velocity amplitude of \hr\ is modulated with
rotation phase.  Such a modulation was predicted by the oblique pulsator
model, and rules out the spotted pulsator model.  However, further analysis
of \halp\ and other lines reveal rotational modulations that cannot easily
be explained using the oblique pulsator model.  In particular, the phase of
the pulsation as measured by the width of the \halp\ line varies with
height in the line.

The variation of the \halp\ bisector shows a very similar pattern to that
observed in \acir, which we have previously attributed to a radial node in
the stellar atmosphere.  However, the striking similarities between the two
stars despite the much shorter period of \acir\ (6.8\,min) argues against
this interpretation unless the structure of the atmosphere is somewhat
different between the two stars.  Alternatively, the bisector variation is
a signature of the degree $\ell$ of the mode and not the overtone value
$n$.

High-resolution studies of the metal lines in roAp stars are needed to
understand fully the form of the pulsation in the atmosphere.
\end{abstract}


\section{Introduction}
\label{sec:hr3831-intro}

The rapidly oscillating Ap (roAp) stars pulsate in high-overtone
non-radial p-modes (\citenb{Kur90}; \citenb{MK95conf}; \citenb{Mat91}, 
\citeyear{Mat97}; \citenb{Shi91}). 
The oscillation spectra of many of these stars reveal frequency
multiplets -- usually triplets -- with equal frequency separations of
the components.  The separations are equal to, or very nearly equal
to, the rotation frequency (determined from the variation of spectrum
and light as the star rotates).

One explanation of the observed triplets is that a set of rotationally
perturbed $m$-modes is excited, e.g., $\ell = 1$ dipole modes with
$m = -1$, 0, +1.  The observed frequencies can be written as
\begin{equation}
 \nu_m = \nu_0 - m (1 - C_{n\ell}) \nu_{\rm rot}
\end{equation}
where $C_{n\ell}$ is a constant that depends on the structure of the
star \cite{Led51}.  For A-star models with the expected pulsation
modes for roAp stars, \citeone{TS95} found 
$C_{n\ell} \approx 0.003$--0.010.  However, in the best-observed roAp
star \hr, \citeone{KKM92} were able to show that 
$C_{n\ell} < 2 \times 10^{-5}$.  This is two orders of magnitude less
than the lowest theoretically predicted values, and suggests that the
frequency splitting is precisely the rotational frequency.

Another explanation for a frequency multiplet is 
a single pulsation mode (e.g., $\ell = 1$, $m = 0$) whose amplitude
is modulated\footnote{In this paper, we use  {\em modulation\/}
  to describe changes in amplitude and phase
  with rotation, and  {\em variation\/} for other types of
  changes.}
by rotation.  This naturally produces a frequency splitting that is
precisely equal to the rotation frequency.  In some roAp stars, the
amplitude of the pulsation has been observed to modulate with rotation
{\em in phase\/} with the magnetic variation.  To explain this
phenomena, the oblique pulsator model was proposed by \citeone{Kur82},
and has been extensively developed by \citename{ST93}
(\citeyear{ST93}; \citenb{TS94}, \citeyear{TS95}).  In this model, the
magnetic and pulsation axes are aligned but are oblique to the
rotation axis.  The observed amplitude modulation is then due to the
variation in the angle between the pulsation axis and the line of
sight as the star rotates.

In an alternative model, the spotted pulsator model \cite{mathys85},
the pulsation and rotation axes are aligned (or there are radial
$\ell = 0$ modes), but the ratio of flux to radius variations varies
over the surface because of differences in the flux and temperature
caused by spots associated with the magnetic field.  The observed
radial-velocity (RV) amplitude is then constant, but the photometric
amplitude modulates as the star rotates.  Therefore, the two models
can be distinguished by RV observations.  \citeone{MWW88} found RV
amplitude variation in HR~1217 which favoured the oblique pulsator
model, but this star is multi-periodic, so the observed variation
could have been caused by beating among pulsation frequencies.
\citeone{KMV94} proposed that the two models could easily be
distinguished by RV measurements of \hr.  This roAp star has one
dominant pulsation mode with a period of 11.7 min, has a large
photometric amplitude modulation including a phase reversal
(\citenb{KMV94}, \citeyear{KVR97}) and is bright enough for accurate
RV measurements ($V=6.25$).

We obtained time-series spectroscopy of \hr\ over one week in 1997
March.  Preliminary analysis of the \halp\ RV measurements was
presented by Baldry, Kurtz \& Bedding (\citeyear{BKB98}; hereafter
BKB) and showed that the radial-velocity amplitude is modulated
with the rotation of the star.  A frequency analysis showed a
frequency triplet with the same spacing and amplitude ratios as
contemporaneous photometric observations.  This is expected in the
oblique pulsator model, and in clear disagreement with the prediction
of the spotted pulsator model.

In this paper we describe more detailed analysis of our \hr\ spectra
using some of the techniques that were used by \citename{BBV98}
(\citeyear{BBV98}, \citeyear{BVB99}; hereafter B98, B99) to
study \acir.  These include cross-correlations of different wavelength
regions, line-shifts of the \halp\ bisector and intensity measurements
across the \halp\ line.  While the oblique pulsator explains most of
the observed rotational modulations, aspects of the spotted pulsator
model and perturbed $m$-modes may also be required to explain our
results.

\subsection{Basic data for \hr}
\label{sec:hr3831-data}

\hr\ (IM~Vel, HD~83368, HIP~47145) has a binary companion with
$V=9.09$ with a separation of 3.29 arcsec.  The visual magnitude of
\hr\ is often quoted as $V=6.17$, but this is from the combined flux
measurement (see \citenb{KMV94}).  Using the correct value of
$V=6.25$, the {\em Hipparcos\/} distance of $72.5 \pm 4$\,pc
\cite{ESA97}, an effective temperature of $8000 \pm 200$\,K
\cite{KMV94} and a bolometric correction of $-0.13$ \cite{Sch82}, we
obtain $L = 13.4 \pm 1.5 L_{\odot}$ and 
$R = 1.9 \pm 0.2 R_{\odot}$.\footnote{In some papers (including
  BKB), the radius of \hr\ is incorrectly quoted as
  2.9$R_{\odot}$.  This may be due to propagation of a typographical
  error.}

The oscillation spectrum of \hr\ has a well-known frequency septuplet
around 1428\mh, plus frequency multiplets at the harmonics
\cite{KVR97}.  However, there are just four frequencies which have
photometric amplitudes above 0.3\,mmag (Johnson $B$) and that are
detectable in our data.  These include a triplet around the principal
mode ($\nu_{-1} = 1423.95$\mh, $\nu_0 = 1428.01$\mh,
$\nu_{+1} = 1432.07$\mh) and the first harmonic ($2\nu_0 = 2856.02$\mh).

The splitting of the triplet ($\nu_{\rm rot} = 4.06$\mh) gives the rotation
period $P_{\rm rot} = 2.851976$\,d, assuming the oblique pulsator model is
correct.
Unlike the triplet in \acir\ \cite{KSM94}, where the amplitudes of the
two side-lobes are about 10 percent of the principal amplitude, the
amplitudes of $\nu_{-1}$ and $\nu_{+1}$ are larger than the amplitude
of the central frequency.  This means that during the rotation cycle
of \hr, the amplitude measured at $\nu_0$ goes through two maxima with
two phase reversals.  In terms of the oblique pulsator model, the star
pulsates in an $\ell = 1$ dipole mode which is sufficiently oblique to
the rotation axis that we see first one pole and
then the other as the star rotates.

Kurtz, Shibahashi \& Goode (\citeyear{KSG90}; \citenb{Kur90}) described
a generalised oblique pulsator model in which the effects of both the
magnetic field and rotation were taken into account (see also
\citenb{DG85}, \citeyear{DG86}).  In this model, the perturbation to
the star's eigenfrequencies by the magnetic field dominates, leading
to the conclusion that the pulsation axis is locked to the magnetic
axis.  Two parameters related to the amplitude ratios in the frequency
triplet were defined:
\begin{equation}
 P_1 = \frac{ A_{+1} + A_{-1} }{ A_0 } = \tan i \tan \beta \: ,
 \label{eqn:p1}
\end{equation}
where $i$ is the inclination of the rotation axis to the line of
sight, and $\beta$ is the angle between the rotation axis and the
pulsation axis; and
\begin{equation}
 P_2 = \frac{ A_{+1} - A_{-1} }{ A_{+1} + A_{-1} } 
 = \frac{ C_{n\ell} \, \nu_{\rm rot} }{ \nu_1^{(1)mag} - \nu_0^{(1)mag} } \: ,
 \label{eqn:p2}
\end{equation}
where the perturbation to the eigenfrequencies (by the magnetic field)
depends on $|m|$ such that $\nu = \nu^{(0)} + \nu_{|m|}^{(1)mag}$.
{}From the measured photometric amplitudes for \hr\ covering 1993 to 1996
\cite{KVR97}, we derive $P_1 = 8.6 \pm 0.2$ and $P_2 = -0.097 \pm 0.004$. 

\section{Observations}

\subsection{Spectroscopy}
\label{sec:hr3831-obs-spec}

We obtained 1400 intermediate-resolution spectra of \hr\ using the
coud\'{e} spectrograph on the 74-inch Telescope at Mt.~Stromlo
Observatory (MSO), Australia.  A log of the observations is shown in
Table~\ref{tab:log97}.
\begin{table}
\caption{Log of the spectroscopic observations of \hr}
\label{tab:log97}
\begin{tabular}{crrc} \hline
UT-date     & No. of & No. of  & Julian dates        \\
            & hours  & spectra & $-$2450000        \\ \hline
            &        &         &                     \\
1997 Mar 10 &  1.37  &   26    &  517.88 -- 517.94   \\
1997 Mar 10 &  3.61  &   38    &  518.12 -- 518.27   \\
1997 Mar 11 & 10.03  &  203    &  518.88 -- 519.29   \\
1997 Mar 12 &  9.59  &  285    &  519.88 -- 520.28   \\
1997 Mar 13 &  2.22  &   59    &  520.88 -- 520.97   \\
1997 Mar 13 &  1.10  &   33    &  521.27 -- 521.32   \\
1997 Mar 14 & 10.57  &  319    &  521.87 -- 522.31   \\
1997 Mar 15 & 10.40  &  262    &  522.88 -- 523.31   \\
1997 Mar 16 &  9.40  &  158    &  523.87 -- 524.27   \\
1997 Mar 17 &  0.53  &   17    &  524.87 -- 524.90   \\ \hline  
\end{tabular}
\end{table}
The wavelength region from 6100\AA\ to 7100\AA\ was observed with a
resolution of about 1.5\AA\ and a dispersion of 0.49\,\AA/pixel.  The
average number of photons/\AA\ in each spectrum was 270\,000, with a
minimum of 60\,000.  The exposure time was 100 seconds,
with an over-head between exposures of 20 seconds.

\begin{figure}
\epsfxsize=9.0cm
\centerline{\epsfbox{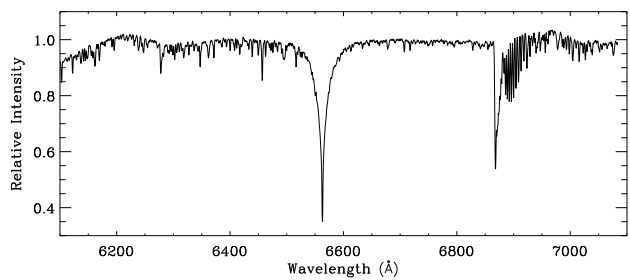}}
\caption{The template spectrum of \hr}
\label{fig:hr3831-spec}
\end{figure}

All CCD images were reduced to spectra using the same
procedures that were used on the Mt.~Stromlo data for B98.  Several
observables were defined from the spectrum (velocities, intensities,
bisector measurements), to produce a time series of measurements for
each observable.  For use in various measurements, a template spectrum
was defined from 25 high-quality spectra
(Fig.~\ref{fig:hr3831-spec}).

\subsection{Photometry}

For the last five years, astronomers at the South African Astronomical
Observatory (SAAO) and the University of Cape Town (UCT) have been
monitoring the oscillations in \hr\ using high-speed photometry
\cite{KVR97}.  For BKB, a subset of their data was selected
which is centred on the time of the spectroscopic observations. These
data were obtained on 26 nights (one hour of high-speed photometry per
night) spanning the dates JD~2450402 to~2450618.

The photometry was analysed to find the amplitudes and phases of the
frequency triplet (see Table~\ref{tab:hr3831-phot}) in order to
compare with the spectroscopic results.
\begin{table}
\caption[Photometric amplitudes and phases of the frequency triplet]{
  Photometric (Johnson $B$) amplitudes and phases$^a$ of the frequency
  triplet (BKB).  {}From this data, $P_1 = 6.9\pm 0.7$ and 
  $P_2 = -0.07\pm 0.02$.}
\label{tab:hr3831-phot}
\begin{tabular}{rrrcc} \hline
frequency & amplitude      & phase\rlap{$^a$}\\
($\mu$Hz) & (mmag)         & (radians) &     \\ \hline
1423.95   & $1.91\pm 0.05$ &  $1.00\pm 0.03$ \\
1428.01   & $0.52\pm 0.05$ & $-1.16\pm 0.10$ \\
1432.07   & $1.66\pm 0.05$ &  $1.03\pm 0.03$ \\ \hline
\end{tabular} \\
$^a$The phases are shifted compared to the those given in BKB due
to a different phase reference point (see Section~\ref{sec:hr3831-tsa}).
\end{table}
The amplitudes of the triplet are within 2$\sigma$ of the amplitudes
from \citeone{KVR97}.  For further details, in particular showing the
pulsation amplitude and phase as a function of rotation phase, see
BKB.

\section{Time-series analysis}
\label{sec:hr3831-tsa}

To illustrate the various methods of analysing the time series for
each spectroscopic observable, we look at an observable which has a
high signal-to-noise ratio in the oscillation spectrum.  This
observable, called \rcw, is a ratio of the \halp\ core to wing
intensity.  In particular, we divided the intensity in a filter with
FWHM$\sim 4$\AA\ by that in a filter with FWHM$\sim 31$\AA.  This is
similar to the \rcw\ measurements used for B99 and BKB but
using slightly narrower filters.

\begin{figure}
\epsfxsize=9.0cm
\centerline{\epsfbox{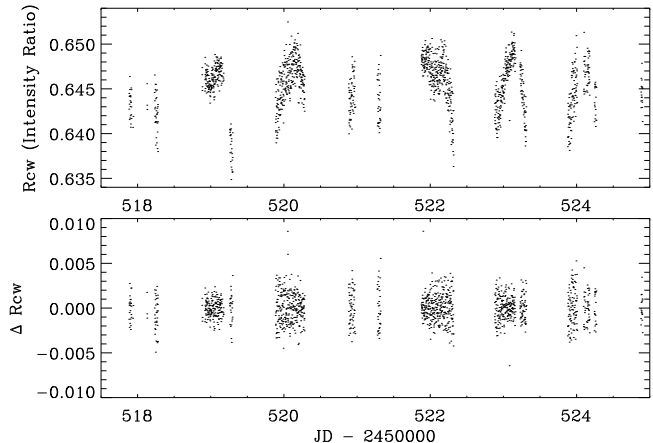}}
\caption{\rcw\ time series before and after high-pass filtering}
\label{fig:rcw-hp}
\end{figure}

Each time series from the spectroscopic measurements was high-pass
filtered and a few outlying points were removed.  The \rcw\ time
series before and after high-pass filtering is shown in
Figure~\ref{fig:rcw-hp}.
There is a variation in the raw data of about 1 percent during the
night.  This is possibly due to varying contamination of the \hr\ 
spectrum from its binary companion.  As the pair of stars rotates in
the coud\'{e} focal plane, there will be less contamination when the
stars line up perpendicular to the slit and more contamination when
the stars line up along the slit.  Alternatively, the variation could
be due to instrumental effects that are a function of elevation of the
telescope.  In either case, the change in the measured oscillation
amplitude will be of the order of 1 percent over the night.  This is
not significant for our results, especially since there is no obvious
systematic effect as a function of the rotation phase of \hr.

Next, a simultaneous fit of the frequency triplet and the first
harmonic was made to the high-pass filtered data, using a weighted
least-squares fitting routine.  The amplitudes and phases were fitted,
using the well-known frequencies (see Section~\ref{sec:hr3831-data}),
by the function $A \sin (2 \pi \nu (t-t_0) + \phi)$ where 
$t_0 =$\,{\footnotesize JD}\,2450522.51746.  Our phase reference 
point $t_0$ is equal to the reference point of \citeone{KVR97} plus 775
times $P_{\rm rot}$.  This maintains the same relationship between the
phases of the frequency triplet.  The fit of the frequency triplet to
the \rcw\ data is shown in Table~\ref{tab:hr3831-rcw} (the results for
three sets of filters are shown).
\begin{table*}
\begin{minipage}{11.5cm}
\caption[\rcw\ amplitudes and phases of the frequency triplet]{
  \rcw\ amplitudes and phases of the frequency triplet in \hr.  The
  observable \rcw\ is the ratio between the mean intensity in a narrow
  filter with FWHM $F1$ and in a filter with FWHM $F2$ centred on
  \halp.  For the three observables in this table, the average values
  are about 0.65 (see Figure~\ref{fig:rcw-hp} for the raw data in the
  second case).}
\label{tab:hr3831-rcw}
\begin{tabular}{ccrrrcc} \hline
$F1$  & $F2$  & frequency & amplitude & phase     & $P_1$ & $P_2$ \\
(\AA) & (\AA) & ($\mu$Hz) & (ppm)     & (radians) &       &     \\ \hline
&& 1423.95& $1340\pm 46$& $-0.82\pm 0.03$& &\\
2.9& 23.5& 
   1428.01& $ 172\pm 45$& $-1.28\pm 0.27$& $15.8\pm 4.7$& $+0.02\pm 0.02$\\
&& 1432.07& $1386\pm 46$& $-0.78\pm 0.03$& &\\ \hline
&& 1423.95& $1334\pm 41$& $-0.71\pm 0.03$& &\\
3.9& 31.4& 
   1428.01& $ 177\pm 40$& $-1.58\pm 0.23$& $14.0\pm 3.5$& $-0.08\pm 0.02$\\
&& 1432.07& $1139\pm 41$& $-0.78\pm 0.04$& &\\ \hline
&& 1423.95& $1107\pm 41$& $-0.64\pm 0.04$& &\\
5.9\rlap{$^a$}& 45.1\rlap{$^a$}& 
   1428.01& $ 163\pm 41$& $-1.70\pm 0.25$& $12.4\pm 3.4$& $-0.10\pm 0.03$\\
&& 1432.07& $ 913\pm 41$& $-0.78\pm 0.05$& &\\ \hline
\end{tabular} \\
$^a$This set of \rcw\ results uses the same filters as the \rcw\ 
results published in BKB.  The results are slightly different due
to a different weighting in the least-squares fitting routine, and
there is also a change in phase reference point.
\end{minipage}
\end{table*}

\begin{figure}
\epsfxsize=9.0cm
\centerline{\epsfbox{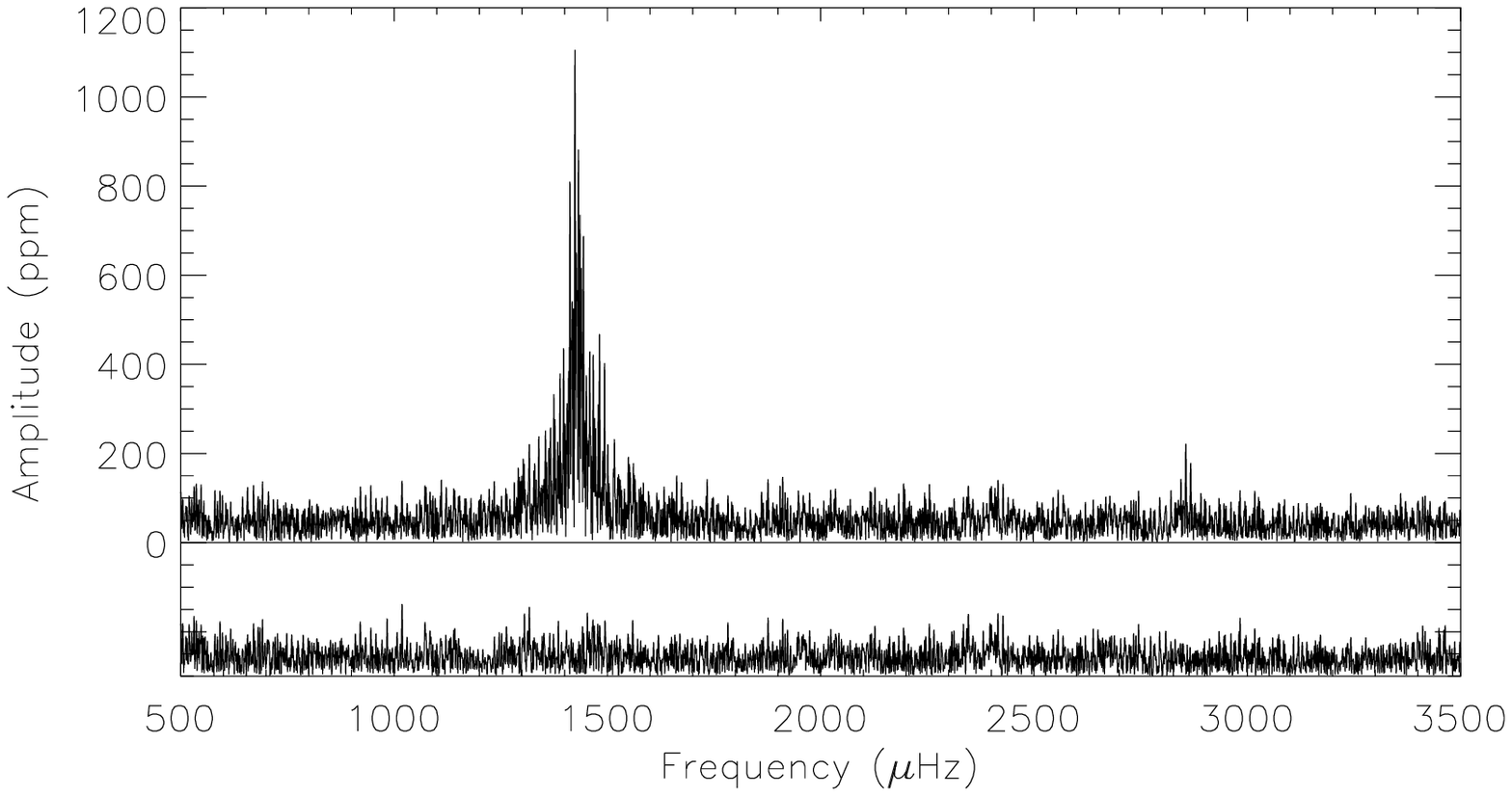}}
\caption[\rcw\ amplitude spectrum]{
  \rcw\ amplitude spectrum. The lower panel is the pre-whitened
  amplitude spectrum after subtracting a simultaneous fit of the
  frequency triplet and the harmonic.}
\label{fig:rcw-sp1}
\vspace{0.3cm}
\epsfxsize=9.0cm
\centerline{\epsfbox{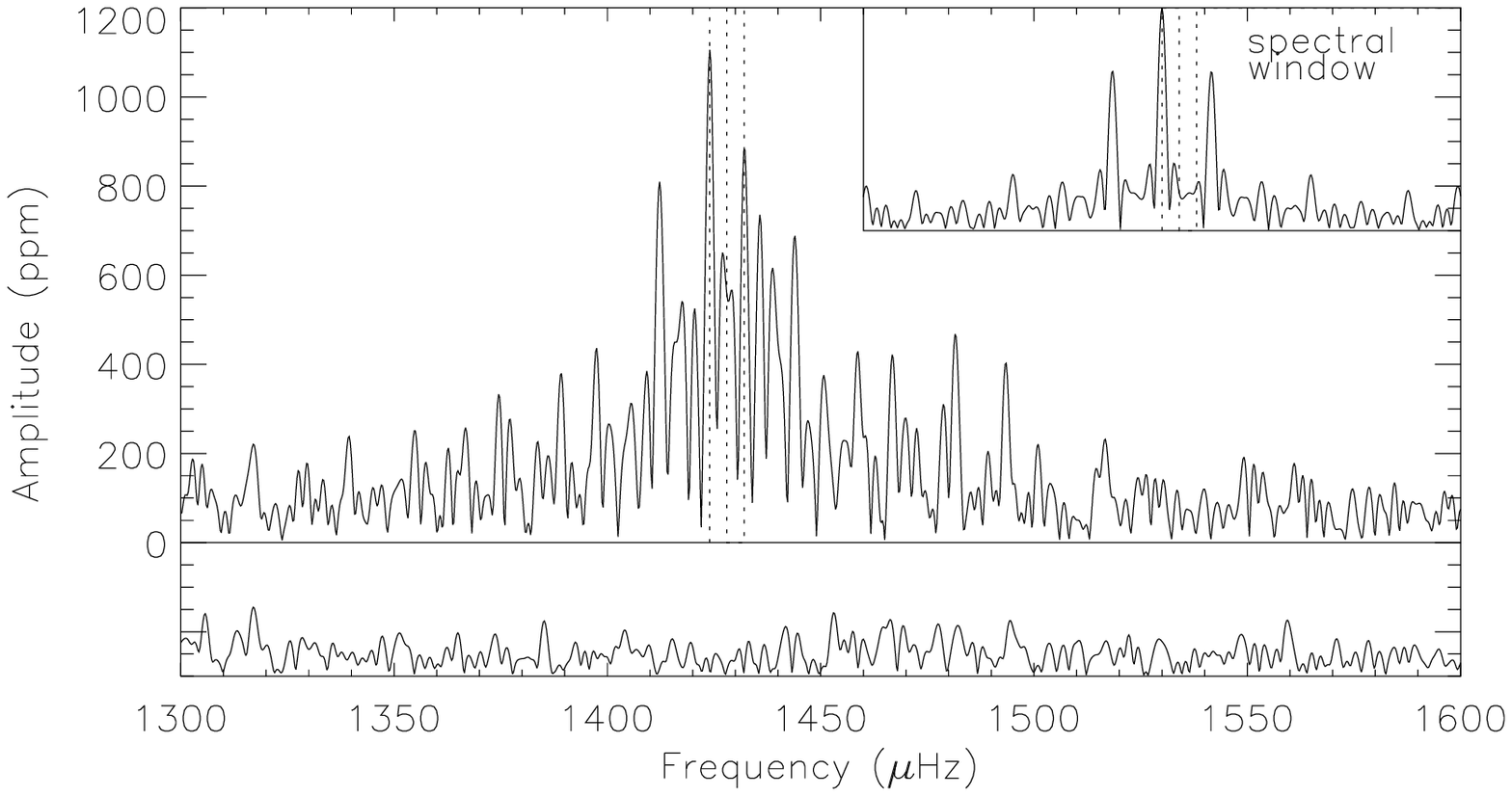}}
\caption[\rcw\ amplitude spectrum]{
  Close-up of Figure~\ref{fig:rcw-sp1}, with the dotted lines showing
  the frequencies of the triplet.  The inset shows the amplitude
  spectrum of the window function on the same frequency scale, with
  the dotted lines having the same spacing as in the triplet.}
\label{fig:rcw-sp2}
\end{figure}

The \rcw\ oscillation amplitude spectrum is shown in 
Figures~\ref{fig:rcw-sp1}--\ref{fig:rcw-sp2}.
Figure~\ref{fig:rcw-sp2} shows that the frequency triplet is resolved,
but with some aliasing or power shifting between the frequencies, as
can be seen from the spectral window.  The amplitude of the spectral
window at $\pm \nu_{\rm rot}$ and at $\pm 2 \nu_{\rm rot}$ is about 20
percent.  Also, the simultaneous fit gives amplitudes in ppm of 1330,
180 and 1140 for the triplet, whereas the amplitudes in the
oscillation spectrum are 1100, 560 and 890 (note that ratio between
$A_{-1}$ and $A_{+1}$ is about the same).

\begin{figure}
\epsfxsize=7.4cm
\centerline{\epsfbox{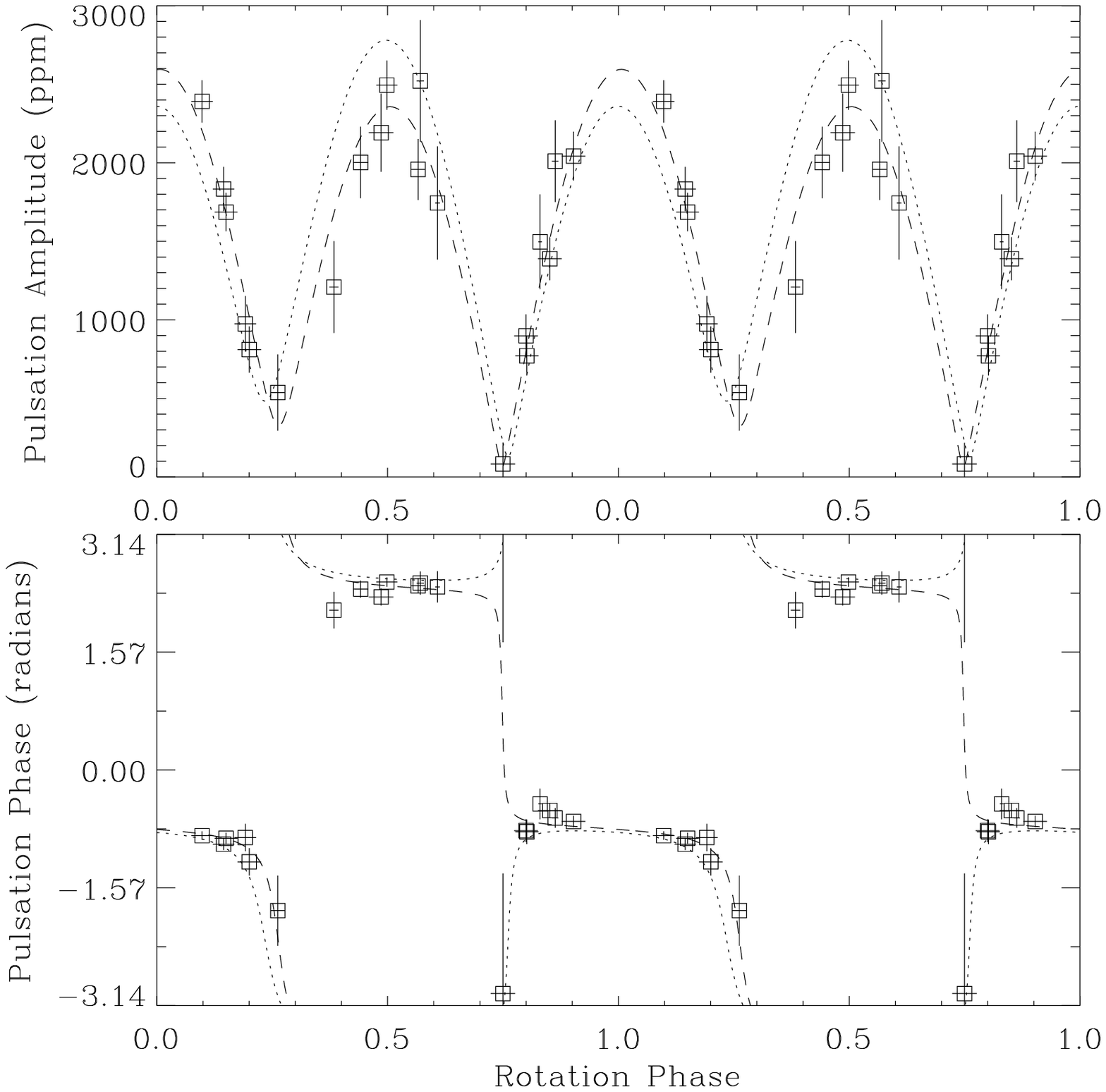}}
\caption[\rcw\ amplitude and phase as a function of rotation phase]{
  \rcw\ amplitude and phase of the central frequency as a function of
  rotation phase.  The squares represent the data divided into 20
  separate time intervals between 0.5 and 3.6 hours long.  The vertical
  lines are error bars, while the horizontal lines show the length of
  the time intervals.  The dashed line represents a fit based on the
  measurement of the frequency triplet from the complete time series,
  and the dotted line represents a fit which is scaled and phase
  shifted from the photometric frequency triplet.  See
  Section~\ref{sec:hr3831-tsa} for further details of the time-series
  analysis.  Note that the data are plotted twice.}
\label{fig:rcw-trip}
\end{figure}

Another way to consider the aliasing is that the data have incomplete
and biased sampling of the rotation phase of \hr.  A good diagnostic
of the data is to plot the amplitude and phase of the pulsation (at
$\nu_0$) as a function of rotation phase.  This is better for
interpretation assuming that there is one mode, that is amplitude
modulated, rather than three distinct modes.  We divided the data into
20 separate time intervals between 0.5 and 3.6 hours long and including
between 17 and 107 spectra.  For each time interval, the amplitude and
phase were measured at 1428.01\mh.  The \rcw\ results are shown in
Figure~\ref{fig:rcw-trip}.
The two lines represent different fits to the data:
\begin{enumerate}
\item The dashed line represents a modulation using the
  amplitudes and phases of the \rcw\ frequency triplet (from
  Table~\ref{tab:hr3831-rcw}), which are from a six parameter fit to
  the complete time series.  In other words, it shows the beating
  effect of the three close frequencies ($\nu_{-1}$, $\nu_0$,
  $\nu_{+1}$).  The good agreement between the modulation in amplitude
  and phase from the separate time intervals and the modulation from the
  frequency triplet supports the accuracy of the simultaneous fit to
  the triplet.
\item The dotted line represents a fit which is obtained by
  scaling and phase shifting from the {\em photometric\/} analysis of the
  triplet (from Table~\ref{tab:hr3831-phot}).  In effect, this is a 
  two-parameter fit to the amplitudes and phases of the separate time
  intervals.  The shape of the amplitude and phase modulation is the
  same as the beating effect of the photometric triplet, while one
  parameter is the scaling of the amplitude and the other parameter is
  the shift of the pulsation phase.  In this case, a reasonable fit is
  obtained to the \rcw\ data.
\end{enumerate}

In conclusion, we have clearly detected the triplet in the
equivalent-width of \halp\ (using \rcw) with the amplitudes of the
three components in about the same ratio as seen in photometry.  The
\rcw\ measurement, as in \acir\ (B99), provides a high
signal-to-noise spectral measurement of the pulsation.

\section{Velocities of wavelength bands}
\label{sec:vel-bands}

In this section, we look at the velocity amplitude and phase of
different wavelength bands using a cross-correlation technique.  The
reduction method is the same as that used on \acir, with a telluric
band used as a velocity fiducial (Sections~3.2--3.3 of B98).  The
spectrum of \hr\ was divided into 90 bands, most having the same
wavelength range as shown in Table~3 of B98, the difference being
that the spectrum of \hr\ was taken from 6100\AA\ to 7100\AA, which
is 100\AA\ higher than the range used in the \acir\ analysis.  The
results for 10 selected bands are shown in
Table~\ref{tab:hr3831-shifts}.
\begin{table*}
\begin{minipage}{13.0cm}
\caption[Velocity amplitude and phases for selected wavelength bands]{
  Velocity amplitude and phases for selected wavelength bands using a
  cross-correlation technique. The bands selected were those having a
  combined signal-to-noise$^a$ ratio of greater than nine for the
  frequency triplet, except only one band across the \halp\ line is
  included (band no.~87).  A telluric band from 6864 to 6881\AA\ is
  used as a velocity reference (band no.~80).}
\label{tab:hr3831-shifts}
\begin{tabular}{crrrccc} \hline
 band\rlap{$^b$} & frequency & amplitude & phase & $P_1$\rlap{$^c$} & $P_2$ 
& Figure \\
range (\AA)      & ($\mu$Hz) & (m\,s$^{-1}$)& (radians) &       &  & \\ \hline
no. 13  & 1423.95& $ 377\pm  51$& $ 0.82\pm 0.13$& &&\\
        & 1428.01& $  92\pm  50$& $-0.67\pm 0.57$& ---& $-0.18\pm 0.12$
& --- \\ 
6140.6--6150.9  & 1432.07& $ 260\pm  51$& $ 0.89\pm 0.20$& &&\\ \hline
no. 14  & 1423.95& $ 524\pm  61$& $ 2.45\pm 0.12$& &&\\
        & 1428.01& $  26\pm  61$&            --- & ---& $-0.32\pm 0.12$
& --- \\ 
6152.9--6164.6  & 1432.07& $ 273\pm  61$& $ 2.41\pm 0.23$& &&\\ \hline
no. 18  & 1423.95& $1108\pm  69$& $ 0.07\pm 0.06$& &&\\
& 1428.01& $ 228\pm  68$& $-1.33\pm 0.30$& $6.6\pm 2.3$& $-0.48\pm 0.07$
& \ref{fig:sA18-trip} \\
6194.0--6197.5  & 1432.07& $ 392\pm  69$& $ 0.09\pm 0.18$& &&\\ \hline
no. 33  & 1423.95& $1140\pm  62$& $ 0.54\pm 0.05$& &&\\
& 1428.01& $ 222\pm  62$& $-0.96\pm 0.28$& $7.6\pm 2.4$& $-0.34\pm 0.05$
& --- \\ 
6325.8--6333.2  & 1432.07& $ 556\pm  62$& $ 0.46\pm 0.11$& &&\\ \hline
no. 42  & 1423.95& $ 600\pm  64$& $ 1.26\pm 0.11$& &&\\
        & 1428.01& $ 118\pm  64$& $-0.77\pm 0.57$& ---& $-0.13\pm 0.09$
& --- \\
6414.0--6422.9  & 1432.07& $ 458\pm  64$& $ 0.70\pm 0.14$& &&\\ \hline
no. 54  & 1423.95& $ 348\pm  60$& $ 1.64\pm 0.17$& &&\\
        & 1428.01& $  88\pm  60$& $-0.44\pm 0.75$& ---& $+0.10\pm 0.11$
& --- \\ 
6521.8--6528.2  & 1432.07& $ 423\pm  60$& $ 1.62\pm 0.14$& &&\\ \hline
no. 58  & 1423.95& $1378\pm 132$& $ 2.36\pm 0.10$& &&\\
        & 1428.01& $ 131\pm 131$& $-0.47\pm 1.49$& ---& $-0.60\pm 0.13$
& --- \\
6596.3--6607.1  & 1432.07& $ 346\pm 132$& $ 2.83\pm 0.39$& &&\\ \hline
no. 81\rlap{$^d$}  & 1423.95& $  63\pm  12$& $-2.59\pm 0.19$& &&\\
        & 1428.01& $  16\pm  12$& $ 1.99\pm 0.80$& ---& $+0.03\pm 0.13$
& --- \\ 
6881.5--6901.6  & 1432.07& $  68\pm  12$& $ 2.54\pm 0.18$& &&\\ \hline
no. 87\rlap{$^e$}  & 1423.95& $ 343\pm  22$& $ 1.58\pm 0.06$& &&\\
& 1428.01& $  94\pm  22$& $-0.81\pm 0.24$& $6.9\pm 1.8$& $-0.05\pm 0.05$
& \ref{fig:sA87-trip} \\
6545.4--6578.2  & 1432.07& $ 309\pm  22$& $ 1.42\pm 0.07$& &&\\ \hline
no. 90\rlap{$^f$}  & 1423.95& $ 743\pm  87$& $ 0.95\pm 0.12$& &&\\
        & 1428.01& $ 116\pm  87$& $-1.15\pm 0.84$& ---& $-0.35\pm 0.12$
& --- \\ 
7070.2--7079.5  & 1432.07& $ 357\pm  87$& $ 1.28\pm 0.25$& &&\\ \hline
\end{tabular} \\
$^a$The rms-noise is about 1.38 times the formal 1$\sigma$ uncertainty on the 
amplitude measurements.
\newline
$^b$The band number relates to the bands used in B98.
\newline
$^c$Only the bands where the central amplitude is higher than 2$\sigma$ 
are included. For the other bands, the uncertainty in this parameter is large 
and significantly non-Gaussian. 
\newline
$^d$This band contains a significant number of telluric lines, 
see discussion in Section~5.1 of B98.
\newline
$^e$This band represents the velocity of the \halp\ line. 
These results are different to those published in BKB because 
the band is wider and the telluric band used as a reference is different.
\newline
$^f$This is a new band not included in B98.
\end{minipage}
\end{table*}

\subsection{\halp\ velocity}
\label{sec:ha-vel-hr3831}

\begin{figure}
\epsfxsize=7.4cm
\centerline{\epsfbox{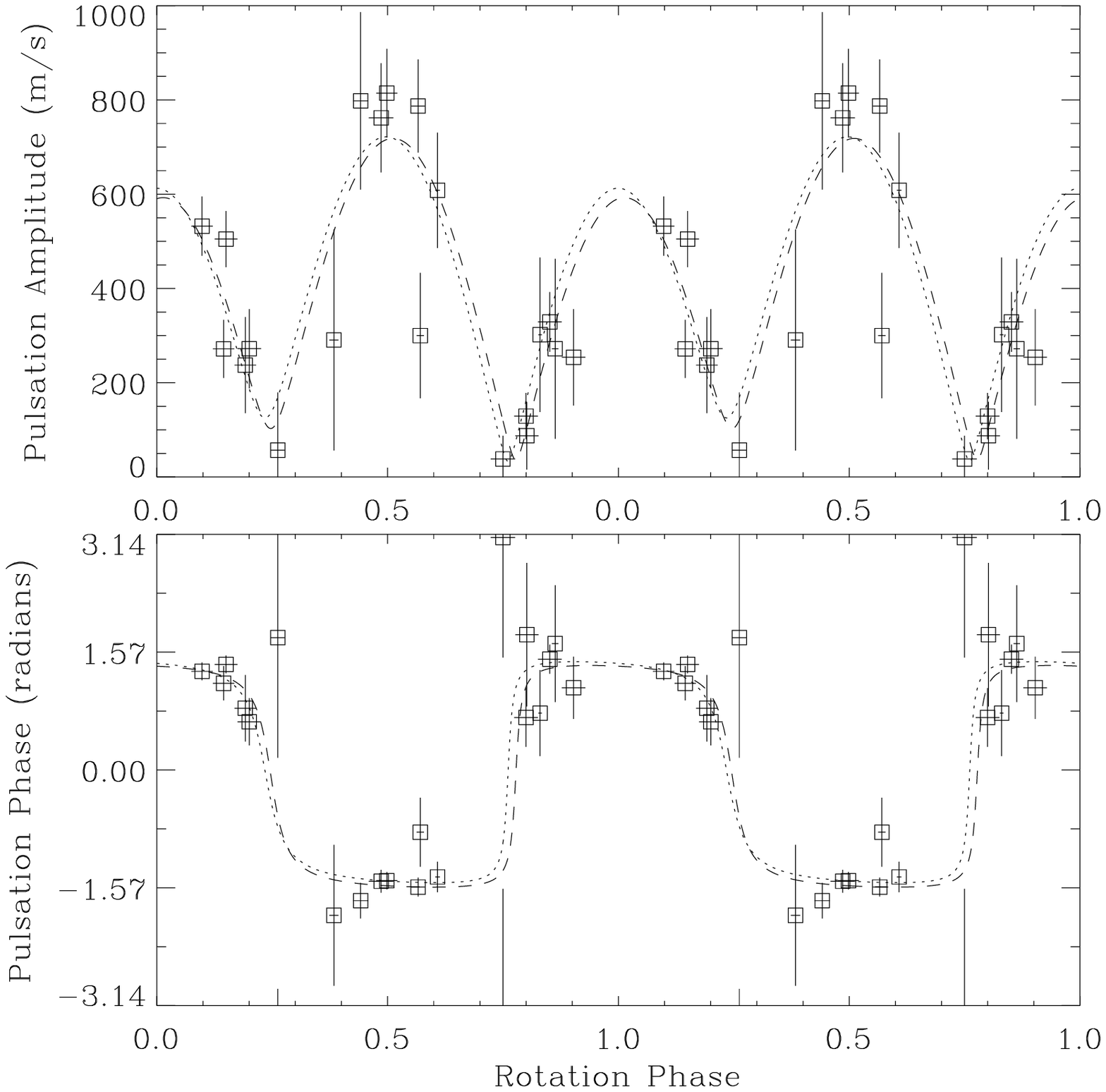}}
\caption[\halp\ velocity amplitude and phase as a function of rotation phase]{
  \halp\ velocity amplitude and phase of the central frequency as a
  function of rotation phase (band no.~87).  Line styles have the same
  meanings as in Figure~\ref{fig:rcw-trip}.  There is excellent
  agreement between the two fits, which rules out the spotted pulsator model.}
\label{fig:sA87-trip}
\end{figure}

First, we look again at the \halp\ velocity (previously analysed for
BKB), as measured using the cross-correlation of band no.~87
(6545--6578\AA).  This band has good signal-to-noise and should not be
significantly affected by blending.  The \halp\ velocity amplitude and
phase as a function of rotation phase are shown in
Figure~\ref{fig:sA87-trip}.
There is excellent agreement between the fit based on the measurement
of the frequency triplet (Table~\ref{tab:hr3831-shifts}) and the fit
scaled and phase shifted from the photometric data
(Table~\ref{tab:hr3831-phot}).  This rules out the spotted pulsator
model. In this model, the pulsation velocity amplitude and phase
should be constant during the rotation of the star, which is clearly
not the case.

The excellent agreement between the \halp\ velocity and photometry, in
terms of the relative modulation as a function of rotation phase, is
reflected in the fact that the parameters $P_1$ and $P_2$ are nearly
the same between the two time series (compare
Table~\ref{tab:hr3831-phot} with band~87 from
Table~\ref{tab:hr3831-shifts}).  Within the oblique pulsator model,
these parameters are expected to be the same for different observables
(see Section~\ref{sec:hr3831-data}).  If the frequency triplet were
caused by three different modes, then the ratios between the
amplitudes of the modes (quantified by $P_1$ and $P_2$) would be
expected to vary depending on the observable.  Therefore, this
agreement between the \halp\ velocity and photometry argues in favour
of the oblique pulsator model rather than for different modes.

However, this is a weak argument because the amplitude ratios could be
nearly the same even in the case of different modes.  The main
arguments against different modes come from the years of photometric
analysis of \hr\ (e.g.: \citenb{KSG90}, \citeyear{KKM93}, \citeyear{KMV94},
\citeyear{KVR97}): (i) the
frequency splitting is exactly or nearly exactly the rotation
frequency (see Section~\ref{sec:hr3831-intro}), and (ii) the ratios
between the amplitudes of the frequency triplet has remained nearly
constant over time.  For these reasons, it is considered unlikely that
the observed frequency triplet in \hr\ is caused by different modes,
in particular rotationally perturbed $m$-modes.

In BKB, a possible rotational phase lag was noted, between the
radial-velocity (RV) amplitude maximum and the photometric amplitude
maximum, of about 0.06 rotation cycles.  However, from the results
shown in Figure~\ref{fig:sA87-trip}, the RV maximum and photometric
maximum are within 0.02 rotation cycles of each other.  The change
from the earlier results arises from the use of a different telluric
reference wavelength region.  In BKB, a large region from 6865 to
6931\AA\ was used, which included bands~80, 81 and 82, in order to
maximise the signal-to-noise ratio.  In this paper, we find a
significant signal arising from the wavelength region of band~81 (see
Table~\ref{tab:hr3831-shifts}) and this is likely to be the cause of
the phase lag noted in BKB.  Band~80 alone is used as the telluric
reference band in this paper because
it has the lowest percentage (about 2\%) of absorption that is
attributable to lines from the star (see discussion in Section~5.1
of B98).  To summarise, the \halp\ RV amplitude is better
represented by the results in this paper, whereas the results in
BKB are slightly contaminated by metal lines around band~81.

\subsection{Metal lines}
\label{sec:hr3831-metal}

\begin{figure}
\epsfxsize=9.0cm
\centerline{\epsfbox{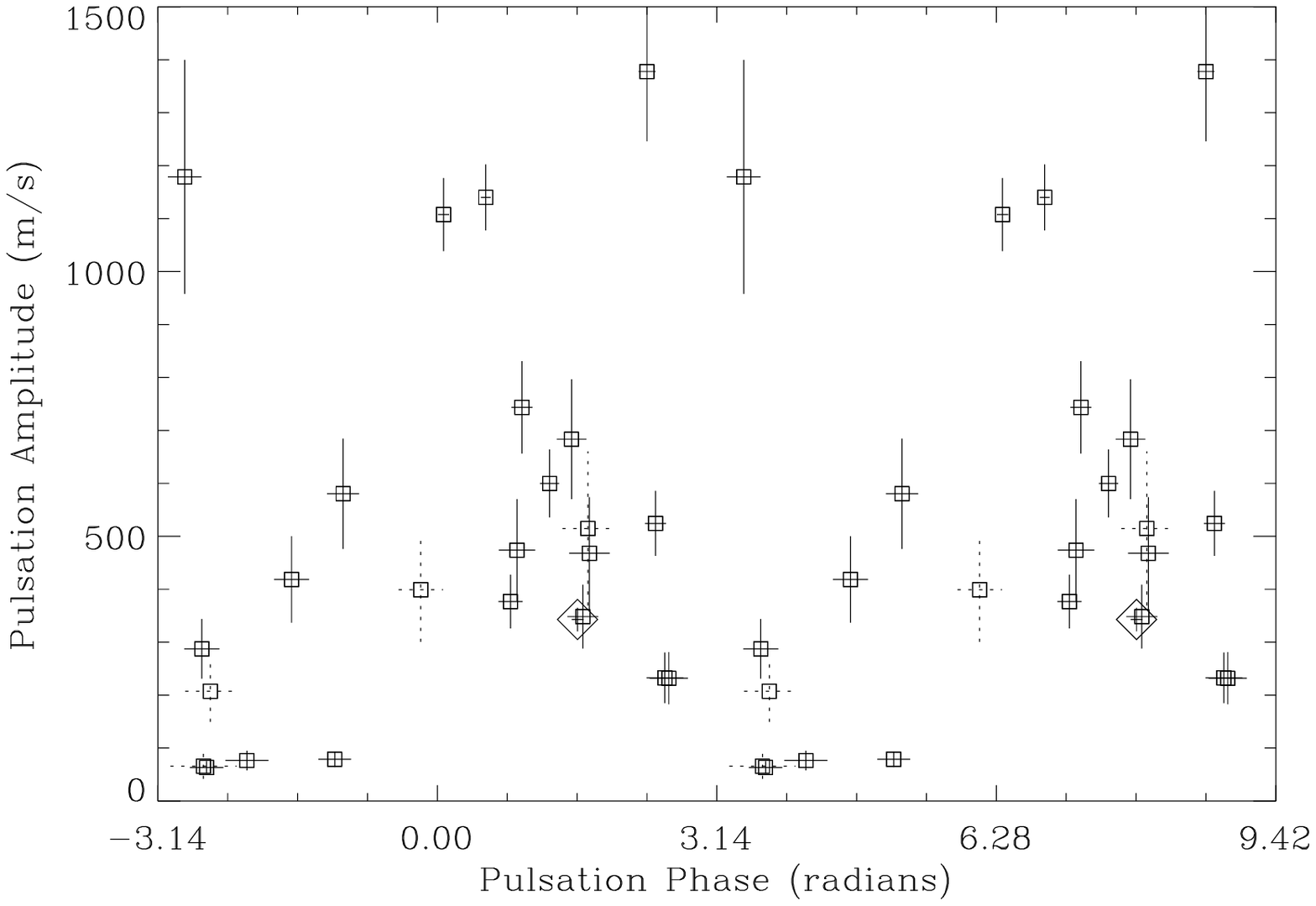}} 
\caption[Velocity amplitudes and phases at the frequency $\nu_{-1}$ in \hr]{
  Velocity amplitudes and phases measured at the frequency $\nu_{-1}$
  in \hr\ for different wavelength bands.  Bands with S/N greater than
  3 are plotted with solid lines, those with lower S/N have dotted
  lines. The same 25 bands are plotted in this figure and in
  Figure~\ref{fig:ampl-ph+1}, chosen so that the {\em combined\/} S/N is
  greater than 5.  Only one band across \halp\ is included and is
  shown using a diamond.  Note that the data are plotted twice for
  clarity.}
\label{fig:ampl-ph-1}
\vspace{0.5cm}
\epsfxsize=9.0cm
\centerline{\epsfbox{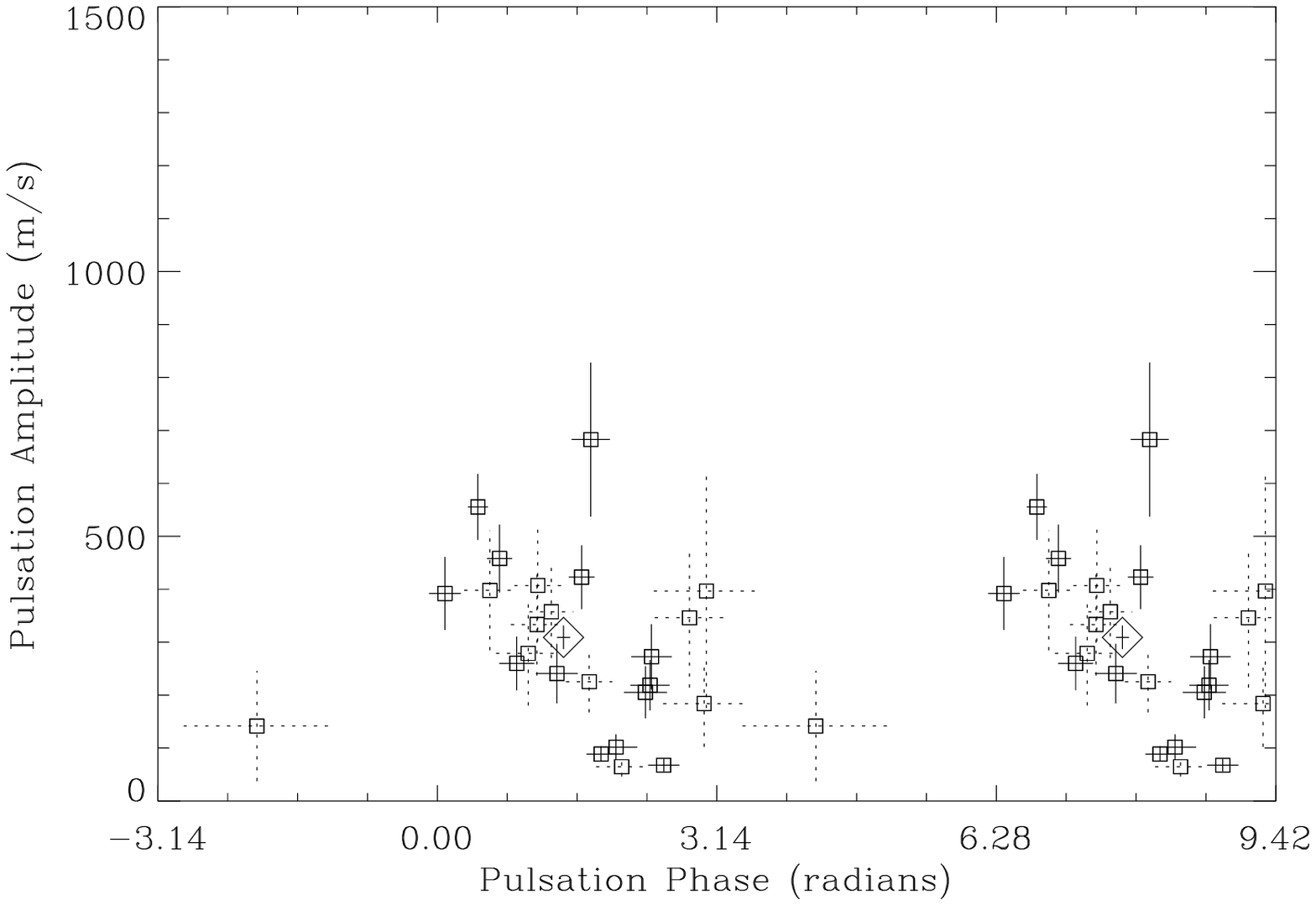}} 
\caption[Velocity amplitudes and phases at the frequency $\nu_{+1}$ in \hr]{
  Same as Figure~\ref{fig:ampl-ph-1}, but for the frequency $\nu_{+1}$.}
\label{fig:ampl-ph+1}
\end{figure}

The velocity amplitude and phase in \hr\ varied significantly between
different wavelength bands, as was seen in \acir\ (B98).
Figures~\ref{fig:ampl-ph-1}--\ref{fig:ampl-ph+1} show the amplitude
versus pulsation phase in \hr\ for the 24 metal bands with the highest 
signal-to-noise ratios (plus one band across \halp).
The results measured at the frequencies $\nu_{-1}$ and $\nu_{+1}$ are
plotted separately, while the amplitude and phase of $\nu_0$ are not
plotted because the signal-to-noise ratio is significantly lower at
this frequency.  There are two aspects to the results which we will
consider: the modulation of the measured pulsation with rotation and
the variation of the amplitude and phase between different wavelength
bands.  The first is characterised by the difference in amplitude and
phase between $\nu_{-1}$ and $\nu_{+1}$.

\subsubsection{Rotational modulation}

There is a significant difference between the results for $\nu_{-1}$
(Fig.~\ref{fig:ampl-ph-1}) and $\nu_{+1}$ (Fig.~\ref{fig:ampl-ph+1}).
Firstly, the amplitudes vary between 0 and 1500\ms\ for $\nu_{-1}$,
whereas the amplitudes for $\nu_{+1}$ are all below 800\ms.  Secondly,
the phases range across 2$\pi$ for $\nu_{-1}$, while the phases for
$\nu_{+1}$ range between 0 and $\pi$ except for one low S/N band.
This is not as expected from the oblique pulsator model.  In this
model, for each band, the phase of $\nu_{-1}$ and $\nu_{+1}$ should be
the same and the amplitudes should be nearly the same 
($A_{+1} \approx 0.87 \times A_{-1}$ from
Table~\ref{tab:hr3831-phot}).  Therefore, these results suggest that
the two frequencies measured are separate modes, rather than
rotational `side lobes' of one principal mode.  However, there is
strong evidence for the oblique pulsator model as described in
Section~\ref{sec:ha-vel-hr3831}.  Therefore, we need to consider
another explanation within the framework of the oblique pulsator
model.  We first look at some selected bands, in terms of their
modulation of amplitude and phase with rotation.

Recall that the oblique pulsator model involves parameters $P_1$ and
$P_2$ (Eqns.~\ref{eqn:p1} \&~\ref{eqn:p2}).  All the measured
spectroscopic observables in this paper produce values for $P_1$ that
are within 3$\sigma$ of the photometric value.  However, there are large
uncertainties
in this parameter because it depends critically on the amplitude
measured at $\nu_0$, which is typically less than two or three times
the noise level in our data.  

\begin{figure}
\epsfxsize=7.4cm
\centerline{\epsfbox{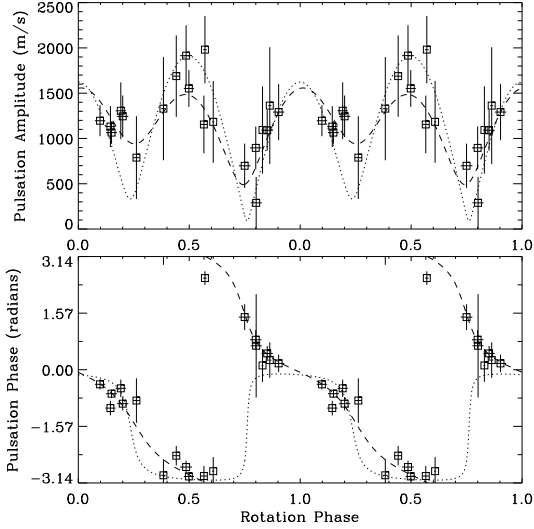}}
\caption[Velocity amplitude and phase of band no.~18]{
  Velocity amplitude and phase of band no.~18 as a function of
  rotation phase.  Line styles have the same
  meanings as in Figure~\ref{fig:rcw-trip}.}
\label{fig:sA18-trip}
\end{figure}

The parameter $P_2$ can be measured with higher accuracy since it does not
depend on the small central amplitude, and we find it does vary
significantly between different observables. 
{}From the measured velocity amplitudes, there are three
spectral bands (18, 33 and 58) that have values of $P_2$ that are
formally 4--5$\sigma$ different from the photometric value (see
Table~\ref{tab:hr3831-shifts}).  To test whether this is a significant
change or just a product of aliasing, we looked at the modulation in
velocity amplitude and phase as a function of rotation phase.
Figure~\ref{fig:sA18-trip} shows this modulation for band~18.
While the amplitude modulation could be fitted reasonably well by
scaling from the photometry (dotted line), the discrepancy is obvious
with the phase modulation.  There is a noticeable pulsation phase
change from rotation phase 0.8 through 1.0 to 0.2, while the
photometric phase is nearly constant.  Similar effects are evident
from the other bands, 33 and~58, for which the $P_2$ value is 
significantly different from the photometric value.

\begin{figure}
\epsfxsize=7.0cm
\centerline{\epsfbox{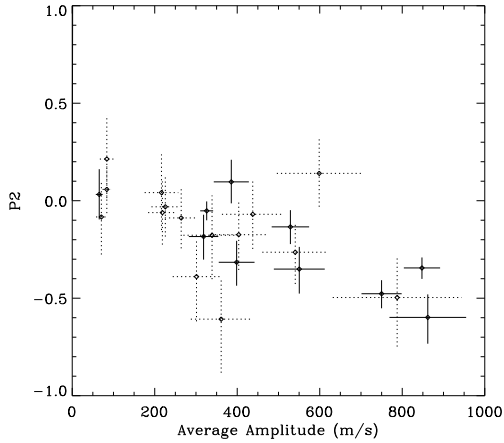}}
\caption{Plot of the parameter $P_2$ versus the average amplitude of 
  frequencies $\nu_{-1}$ and $\nu_{+1}$, with 1$\sigma$-error bars,
  for the 25 bands that are shown in Figures~\ref{fig:ampl-ph-1}
  and~\ref{fig:ampl-ph+1}.  Points with solid lines represent the
  higher S/N bands given in Table~\ref{tab:hr3831-shifts}.  Note that
  the four lowest amplitude bands in this figure are contaminated by
  telluric lines. For these bands, the cross-correlation measurement
  of the velocity amplitude is reduced compared to an uncontaminated
  band (with equivalent stellar behaviour).}
\label{fig:p2-ampl}
\end{figure}

Figure~\ref{fig:p2-ampl} shows a plot of $P_2$ versus the average 
amplitude of frequencies $\nu_{-1}$ and $\nu_{+1}$. 
There is a noticeable tendency for higher amplitude bands to have lower
values of $P_2$.  This could be related to the line-formation depth or
position on the surface (i.e., spots).
Alternatively, a systematic effect (e.g., blending) 
could be increasing the measured amplitude at $\nu_{-1}$ only 
causing an increase in average amplitude and a decrease in $P_2$. 

\begin{figure}
\epsfxsize=7.4cm
\centerline{\epsfbox{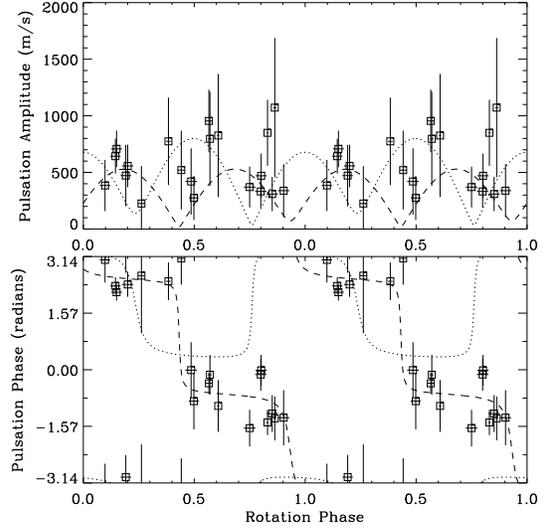}}
\caption{Velocity amplitude and phase of band no.~48 as a function of
  rotation phase.  Line styles have the same
  meanings as in Figure~\ref{fig:rcw-trip}.  }
\label{fig:sA48-trip}
\end{figure}

As well as the parameters $P_1$ and $P_2$, which relate the amplitudes
of the frequency triplet, there are also phase differences between the
frequencies.  For many bands, there is significant difference between
the phases measured at $\nu_{-1}$ and $\nu_{+1}$.  In our
measurements, the equality of phase $\phi_{-1}$ and $\phi_{+1}$ means
that the amplitude maxima occur at rotation phases 0.0 and 0.5 with
phase jumps in between.  Therefore, any phase difference, 
$\phi_d = \phi_{+1} - \phi_{-1}$, implies that there is a shift of the
amplitude maxima.  For example, there is a small shift in the
amplitude maxima of band~42 ($\phi_d = -0.56 \pm 0.18$) and larger
shifts, but with lower S/N in the amplitude spectra, of bands~48
($\phi_d = -2.29 \pm 0.31$) and~82 ($\phi_d = 2.99 \pm 0.25$).
Figure~\ref{fig:sA48-trip} shows the modulation for band~48. 
Note that the pulsation-phase jumps occur out of (rotation) phase 
with the photometric pulsation-phase jumps. 

These results are not sufficient evidence to discard the oblique
pulsator model, because they could be explained by variations in the
ratios between the metal lines in the bands, due to spots, as the star
rotates.  Alternatively, there is a possibility that there is an
actual change in the pulsation phase and amplitude (as opposed to a
pseudo-change caused by a variation in blending) associated with spots
on the star.  So while the spotted pulsator model is ruled out in its
original form, spots may be having an influence on the pulsation.

\subsubsection{Amplitude and phase variations between bands}

In B98, it was suggested that the amplitude and phase variations
between wavelength bands in \acir\ could be explained by a radial node
in the atmosphere.  We see amplitude and phase variations in \hr\ that
are somewhat similar in the case of $\nu_{-1}$
(Fig.~\ref{fig:ampl-ph-1}), and with smaller but still significant
variations in the case of $\nu_{+1}$ (Fig.~\ref{fig:ampl-ph+1}).
However, there is no obvious division between two sets of bands that
are pulsating in anti-phase with each other, as was seen in \acir\ 
(B98).  If these \hr\ results represent true velocity amplitudes
and phases at various depths in the atmosphere, then an explanation
that includes running waves is necessary.  Alternative explanations
could involve spots on the surface or blending, which causes the
measured phase to deviate from the true velocity phase (for further
discussion see Section~\ref{sec:depth-surf}).

\section{H$\balpha$ profile variations}
\label{sec:hr3831-bis-ha}

\begin{figure}
\epsfxsize=7.0cm
\centerline{\epsfbox{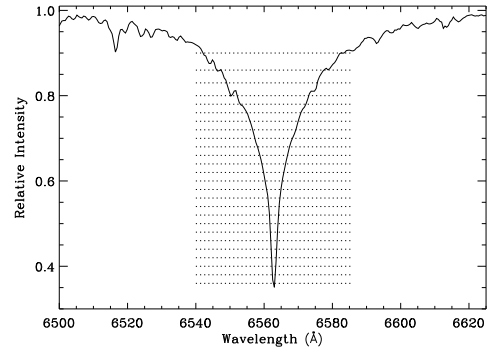}}
\caption[The \halp\ line in \hr]{
  The \halp\ line in \hr.  The dotted lines divide the 25 contiguous
  sections used in the bisector-velocity and width analysis
  (Section~\ref{sec:hr3831-bis-ha}).}
\label{fig:hr3831-sp-ha}
\end{figure}

In \acir\ we found intriguing behaviour in the \halp\ line that implied the
existence of a radial node in the atmosphere (B99).  We now describe a
similar analysis for \hr.
The \halp\ line in each spectrum was divided into 25 contiguous
horizontal sections, between relative intensities 0.35 and 0.9
(Fig.~\ref{fig:hr3831-sp-ha}).
For each section, a bisector velocity (average position of the two
sides) and a width were calculated.  For the velocity measurements, a
telluric band (no.~80) was used as a fiducial.

\subsection{Bisector velocities}

\begin{figure}
\epsfxsize=8.5cm
\centerline{\epsfbox{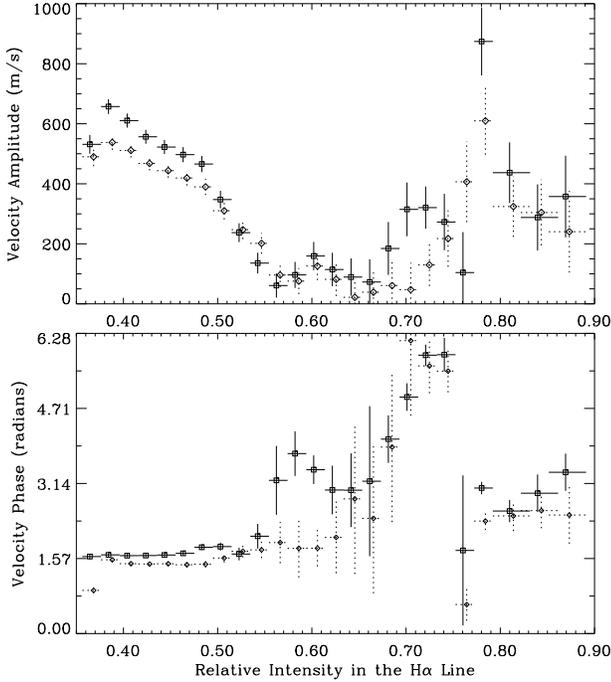}}
\caption[Amplitudes and phases of the pulsation for the bisector velocity]{
  Amplitudes and phases of the pulsation for the bisector velocity at
  different heights in the \halp\ line.  Squares with solid lines
  represent $\nu_{-1}$ and diamonds with dotted lines represent
  $\nu_{+1}$.  For each measurement, the vertical line is an error bar
  while the horizontal line shows the extent of the section in the
  \halp\ line.}
\label{fig:hr3831-bis-vel}
\end{figure}

\begin{figure}
\epsfxsize=7.4cm
\centerline{\epsfbox{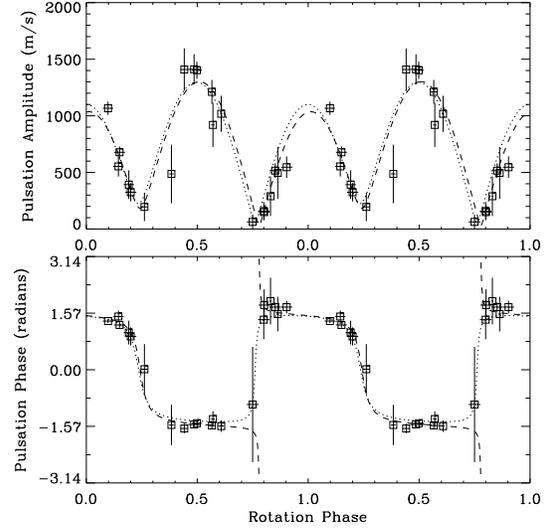}}
\caption[Velocity amplitude and phase of the \halp\ bisector]{
  Velocity amplitude and phase of the \halp\ bisector at height 0.40
  as a function of rotation phase.  Line styles have the same
  meanings as in Figure~\ref{fig:rcw-trip}.  }
\label{fig:bB03-trip}
\end{figure}

The velocity amplitude and phase of the pulsation as a function of
relative intensity in the \halp\ line are shown in
Figure~\ref{fig:hr3831-bis-vel}, with measurements at frequencies
$\nu_{-1}$ and $\nu_{+1}$.
The amplitudes and phases are in good agreement between the two
frequencies, which means that the shape of the rotational modulation
is similar for each velocity measurement.  Figure~\ref{fig:bB03-trip}
shows the rotational modulation from the velocity at a height of 0.40.
It is similar to the modulation from the cross-correlation velocity of
\halp\ (Fig.~\ref{fig:sA87-trip}), but with about twice the
amplitude.  This is expected because the cross-correlation velocity is
a weighted average of the bisector velocities and the velocity
amplitude at 0.40 is higher than average.

\begin{figure}
\epsfxsize=8.5cm
\centerline{\epsfbox{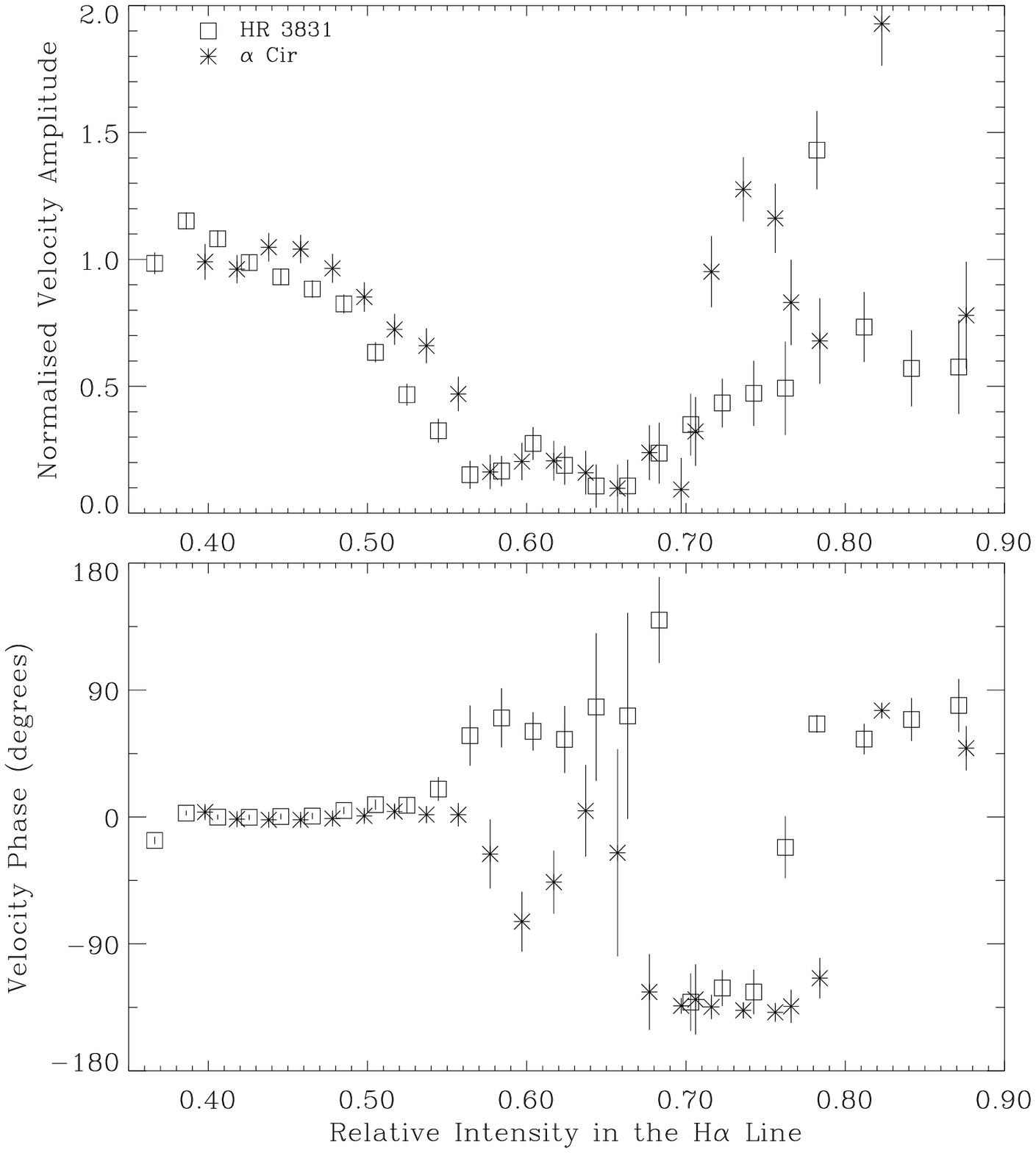}}
\caption[Comparison of the \halp\ bisector variations]{
  Comparison of the \halp\ bisector variations between \hr\ and \acir.
  This figure is essentially a normalised overplot of
  Figure~\ref{fig:hr3831-bis-vel} of this paper and Figure~7 of B98,
  except that for \hr\ the measurements for $\nu_{-1}$ and $\nu_{+1}$
  were averaged and for \acir\ the Mt.~Stromlo and La~Silla data were
  combined.  Additionally, the amplitudes were normalised by dividing
  by an average taken between heights 0.40 and 0.45 (519\ms\ for \hr\ 
  and 267\ms\ for \acir) and the phases were shifted.}
\label{fig:bis-compare} 
\end{figure}

The variation in amplitude and phase as a function of relative
intensity (Fig.~\ref{fig:hr3831-bis-vel}) is similar to that seen in
\acir, particularly the drop in amplitude between heights 0.4 and 0.6
(a comparison is shown in Figure~\ref{fig:bis-compare}).
However, assuming that the bisector line-shift represents the velocity
at different heights in the atmosphere, we expect to see less relative
variation in the bisector-velocity amplitude in \hr\ than \acir.  This
is because the separation between radial nodes in the atmosphere is
larger in \hr\ due to the fact that it pulsates at a lower frequency
and that the two stars have similar fundamental properties.

Why are the behaviours of \acir\ and \hr\ so similar?  Two possible
explanations are: (i)~the \halp\ line forming region in \hr\ is more
extended than in \acir\ and coincidently the \halp\ bisector behaves in a
similar way; (ii)~the assumption that the bisector represents velocities at
different heights is incorrect.

The variation in amplitude and phase is less similar to \acir\ above a
height of 0.7.  For instance, there is a noticeable amplitude peak and
phase reversal around 0.73 in the bisector of \acir\ which is not
evident in \hr.  The bisector measurements above a height of 0.7 in
\acir\ are partially affected by blending (Section~3.2 of B99).
Therefore, some of the differences in the bisector-velocity results
between the two stars are probably due to a difference in blending.

To summarise, there is no significant difference in the rotational
modulation of the bisector velocity between various heights.  For the
measurements below a height of 0.7, the relative variation in
amplitude and phase is very similar to that observed in \acir, which
is puzzling.  At each height, the amplitudes at the two frequencies in
\hr\ are about twice the amplitude of the principal frequency in
\acir.

\subsection{Widths}

\begin{figure}
\epsfxsize=8.5cm
\centerline{\epsfbox{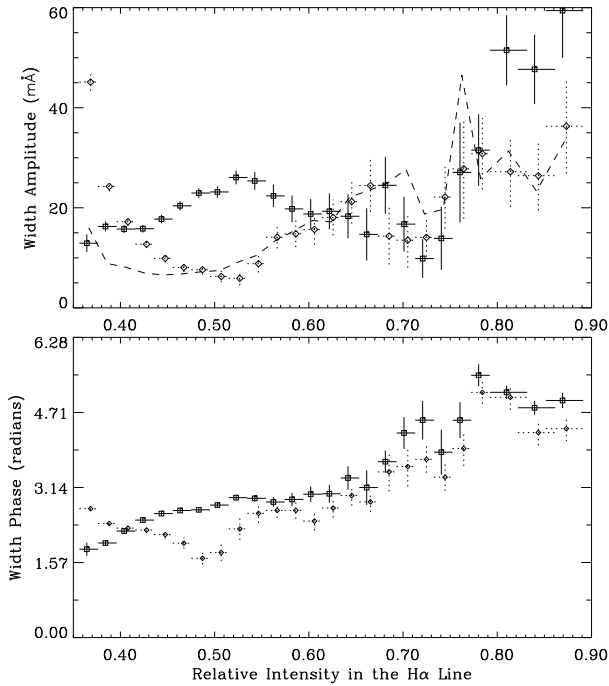}}
\caption[Amplitudes and phases of the pulsation for the width]{
  Amplitudes and phases of the pulsation for the width at different
  heights in the \halp\ line.  Points with solid lines represent
  $\nu_{-1}$ and points with dotted lines represent the $\nu_{+1}$.
  For each measurement, the vertical line is an error bar while the
  horizontal line shows the extent of the section in the \halp\ line.
  The dashed line shows the effect on the width amplitude of an
  oscillation with an EW amplitude of 1000\,ppm, with the profile
  variation described in Section~4.2 of B99.  Note that 10 m\AA\ 
  is equivalent to a velocity of 457\ms\ at \halp.} 
\label{fig:hr3831-bis-wid}
\end{figure}

Changes in the width of the \halp\ line in \hr\ at various heights are
shown in Figure~\ref{fig:hr3831-bis-wid}.
The width amplitude and phase represent oscillations about a mean
width, with measurements at two frequencies.  These measurements are
related to intensity and EW measurements, with the line resolved
vertically rather than horizontally.  The advantage of width
measurements is that the width is naturally independent of the
bisector velocity, i.e., the width is the difference in position
between two sides of a line whereas the velocity is the average
position.  In order to compare with EW measurements, the dashed line
shows the simulated width amplitudes for the profile variation
described in Section~4.2 of B99 (see also Figure~6 of that paper)
with an EW amplitude of 1000\,ppm.

\begin{figure}
\epsfxsize=7.4cm
\centerline{\epsfbox{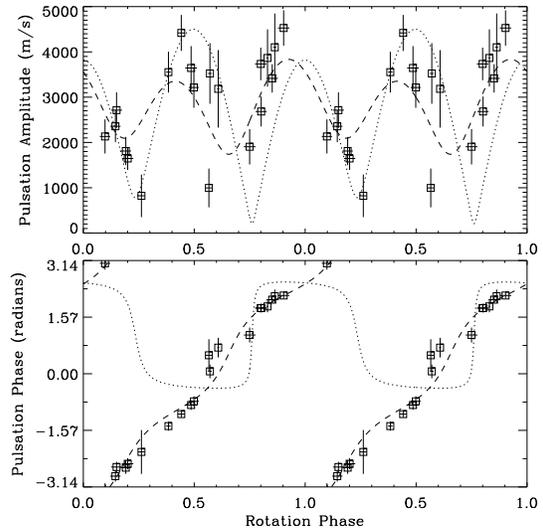}}
\caption[]{Width amplitude and phase of \halp\ at height 0.36 as a function
  of rotation phase.  Line styles have the same
  meanings as in Figure~\ref{fig:rcw-trip}.  }
\label{fig:bB12-trip}
\end{figure}

\begin{figure}
\epsfxsize=7.4cm
\centerline{\epsfbox{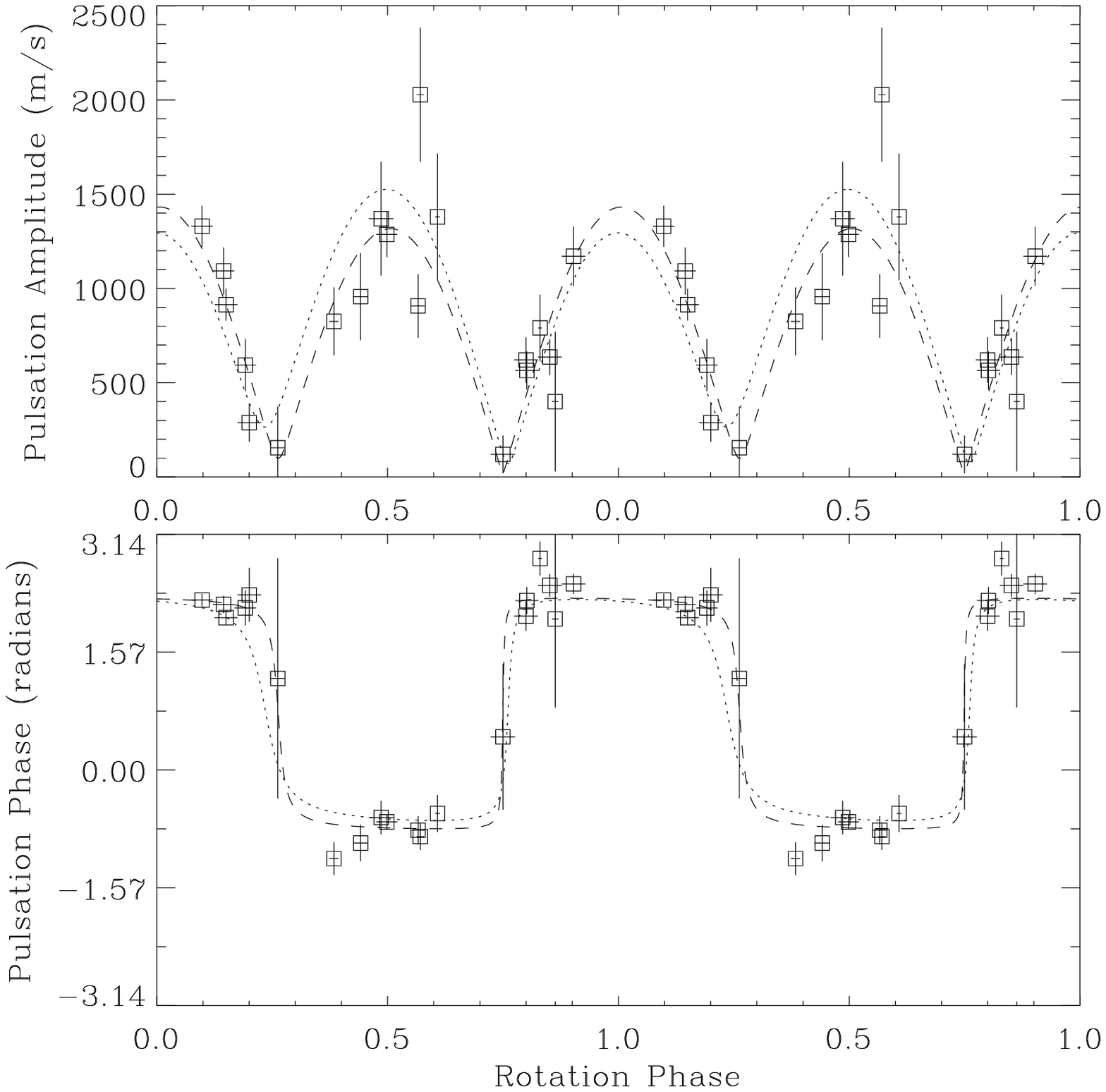}}
\caption{Width amplitude and phase of \halp\ at height 0.42 as a function of
  rotation phase.  Line styles have the same
  meanings as in Figure~\ref{fig:rcw-trip}.  }
\label{fig:bB16-trip}
\end{figure}

\begin{figure}
\epsfxsize=7.4cm
\centerline{\epsfbox{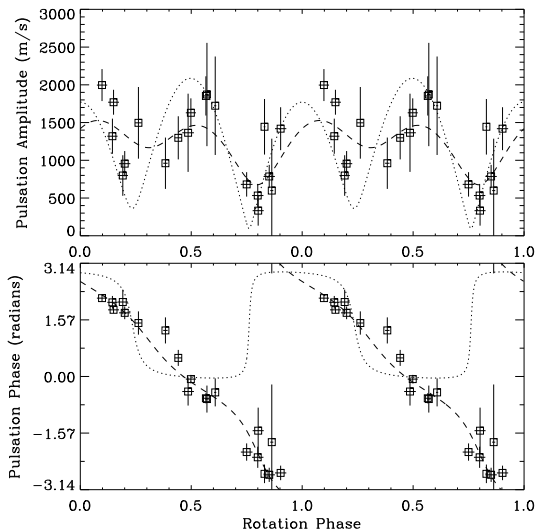}}
\caption{Width amplitude and phase of \halp\ at height 0.53 as a function of
  rotation phase.  Line styles have the same
  meanings as in Figure~\ref{fig:rcw-trip}.  }
\label{fig:bA31-trip}
\end{figure}

The most striking feature of Figure~\ref{fig:hr3831-bis-wid} is the
difference in the measurements between the two
frequencies.  This means that, unlike for the bisector velocities, the
shape of the rotational modulation is varying between different
heights.  We show three examples:
\begin{enumerate}
\item Figure~\ref{fig:bB12-trip} shows the modulation in width
  amplitude and phase at a height of 0.36.  The pulsation width phase
  increases steadily during the rotation and $P_2$ is large ($0.54 \pm 0.05$).
\item Figure~\ref{fig:bB16-trip} shows the modulation at height
  0.42.  There is good agreement with the fit derived from the
  photometric triplet and $P_2$ is $-0.03 \pm 0.04$, which is close to the
  photometric value.
\item Figure~\ref{fig:bA31-trip} shows the modulation at
  height 0.53.  The width phase decreases steadily during the rotation
  and $P_2$ is $-0.63 \pm 0.07$.
\end{enumerate}
To check the lowest measurement at height 0.36, which is sensitive to
intensity changes in the core of the line, we took an intensity
measurement using a filter with a full-width half-maximum of about
1\AA.  This measurement showed a similar rotational modulation, with
the parameter $P_2 = 0.33 \pm 0.03$.  

Something unexpected is certainly happening here.  The \halp\ width changes
at the two extremes, heights 0.36 and 0.53, are in phase at rotation phases
around 0.5 and 1.0, but in anti-phase at phases around 0.25 and 0.75.  At
both heights, the relative amplitude modulation (as defined by the dashed
line) is less than for other observables.

Looking at Figure~\ref{fig:hr3831-bis-wid}, the simple explanation is that
we are seeing two modes with different \halp\ profile variations.  An
alternative explanation -- if \hr\ is pulsating in an oblique mode -- is
that, as the star rotates, the profile variation modulates due to viewing
different aspects of the same mode.

\section{Discussion and conclusions}

\subsection{Velocity amplitudes and phases} 

\subsubsection{Another case: \gequ}
\label{sec:gamma-equ}

\citeone{KH98} observed radial-velocity variations in the roAp star
\gequ.  Their observations consisted of high-resolution spectra in the
range 5000--6000\AA, covering 3.5\,h on 1994 September 13.  There are
four modes in the range 1330--1430\mh\ \cite{MWN96}, which are
unresolved in that data set.  \citename{KH98} fitted a sine curve of
fixed frequency (1380\mh), but variable amplitude and phase, to the
velocity measurements of each individual line, effectively treating
the modes as a single pulsation.  They discovered that the velocity
amplitude varied significantly from line to line (see B98 and
Section~\ref{sec:vel-bands} for similar results on \acir\ and \hr).

\begin{figure}
\epsfxsize=9.0cm
\centerline{\epsfbox{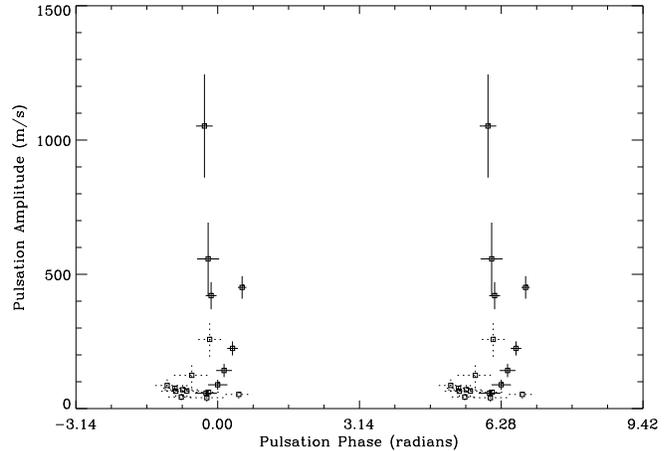}} 
\caption[Amplitudes and phases of the pulsation in \gequ]{
  Amplitudes and phases of the pulsation in \gequ\ for different metal
  lines, using results from \citeone{KH98}.  Lines with $A/\sigma_a$
  greater than 4 are plotted with solid lines, those with 
  $2.5 < A/\sigma_a < 4$ are plotted with dotted lines.  Note that the
  data are plotted twice for comparison with Figures~7 and~10 of
  B98 and Figures~\ref{fig:ampl-ph-1} and~\ref{fig:ampl-ph+1} of
  this paper.}
\label{fig:kh-amp-ph}
\end{figure}

We analysed some of these results on \gequ, given in Table~1 of
\citeone{KH98}, in order to compare with the results on \acir\ and
\hr.  {}From their table, which lists 70 metal lines, we took the
amplitude $A$ (m/s), $\sigma_a$ (m/s) and the `Time of Maximum' (JD).
Note that the times of maximum quoted by \citename{KH98} cover several
pulsation cycles.  We converted the times to phases using the
pulsation period of 0.008387\,d, and the phases were shifted so that a
weighted mean was approximately zero.  The phase uncertainty was taken to be
the arcsin\,($\sigma_a / A$).  This phase uncertainty is in good agreement
with the $\sigma_b$ (days) quoted in their table for lines where 
$A / \sigma_a$ is greater than 2.  Figure~\ref{fig:kh-amp-ph} shows
amplitude versus phase for the 19 lines with the highest
signal-to-noise ratio.
The higher amplitude bands, plotted with solid lines, cover a range of
0.8\,radians in pulsation phase ($-0.26$ to $+0.54$ in the figure).
This is significantly less than the variation in phase seen in \acir\ 
(Fig.~7 of B98) and \hr\ (Fig.~\ref{fig:ampl-ph-1}), where the
bands cover virtually the whole range of phases ($2\pi$).  Note that
the rms-noise level in an oscillation spectrum is about 
$1.38 \sigma_a$, therefore, $A / \sigma > 4$ is comparable to 
$A / {\rm noise} > 3$.

The difference between the results for \gequ\ (\citename{KH98}) and
those for the other two roAp stars, in terms of the variation in
phase between different metal bands, could be due to: (i) wavelength
region studied, 5000--6000 vs.\ 6000--7000, (ii) spectrograph
resolution, 0.2\AA\ vs.\ 1.5\AA, (iii) pulsational characteristics of
the stars.  Further high-resolution studies of these roAp stars will be
able to distinguish among these causes.

\subsubsection{Depth and surface effects}
\label{sec:depth-surf}

What are the causes of the velocity amplitude and phase variations,
and of the \halp\ bisector variations?

The main hypothesis is that there is a significant change in pulsation
amplitude and phase with geometric depth in the atmosphere of an roAp
star, with a radial node of a standing wave situated in the observable
atmosphere.  Absorption lines are formed at different depths and
therefore sample different parts of the standing wave.  This can
explain a complete range of observed amplitudes plus a phase reversal.

An alternative hypothesis for these variations is that the amplitude
and phase of a non-radial pulsation mode will vary over the surface of
a star, and that each spectroscopic observable is related to an
integral over the surface.  Therefore, if the integrals are
sufficiently different between observables, the measured amplitude and
phase may vary significantly.  These variations could be related to
limb-darkening and\,/\,or spots.

Neither hypothesis can fully explain the range of phases seen in \acir\ and
\hr.  Blending effects may cause deviations from the true velocity phases
that are in addition to any atmospheric depth and surface effects.  The
smaller phase variations seen in the high-resolution results of \gequ\ lend
some support to this.  Blending can be considered as blurring the
distinction between velocity and temperature changes in a star.

Surface effects would be a good explanation if roAp stars were
pulsating in modes with $\ell \ge 3$ because, for these modes, there
are several patches on the surface that have alternating phase.
However, the frequency triplets in many roAp stars suggest that dipole
modes (i.e., $\ell = 1$) are predominant.  For these modes, the measured
velocity amplitude and phase are not expected to vary significantly
between different observables.  For example, for a
dipole mode with an axis that is aligned near to our line of sight,
the pulsation phase will be the same across most of the viewed
surface.

The main arguments for depth effects being the principal cause of the
amplitude and phase variations are: (i)~a radial node is plausibly
situated in the observable atmosphere (Gautschy, Saio \& Harzenmoser
\citeyear{GSH98}), and (ii)~the bisector velocity varies with
height in the \halp\ line.  In Section~3.3 of B99, the possibility
of systematic errors causing the bisector-velocity variations was
discussed.  Here, we make a simple calculation to test the accuracy of
these velocities below a relative intensity of 0.7 in the line.

For \acir\ and \hr\ ($\nu_{-1}$ and $\nu_{+1}$), we averaged the
bisector-velocity amplitudes below a height of 0.7 (giving half weight
to the amplitudes between 0.6 and 0.7) in order to compare with the
cross-correlation measurements.  This is a plausible estimate for the
cross-correlation because the slope of the \halp\ profile is steep and
nearly constant between 0.4 and 0.6, and is less steep above 0.6.  For
\acir, we obtained an estimate of $\sim$170\ms\ while the actual
cross-correlation measurements gave 168--182\ms\ (depending on the
band, nos.\ 85--88).  For \hr, we obtained estimates of $\sim$340 and
$\sim$290\ms\ for the two frequencies while the actual measurements
gave 338--361 and 296--318\ms\ respectively.  The good agreement,
between the estimates obtained from the bisector analysis and the
velocity amplitudes from the cross-correlation measurements (using
four different bands), argues that the bisector velocities are
accurate.

However, there is a similarity between the \acir\ and \hr\ bisector
results (see Figure~\ref{fig:bis-compare}) that is possibly
inconsistent with depth effects being the cause of the bisector
variations.  This is because the pulsations of these two stars have
significantly different periods (7 and 12 minutes) and, therefore,
have different vertical wavelengths, assuming that the sound speed is
approximately the same between their atmospheres.  If depth effects
are causing the bulk of the observed bisector variations, the \halp\ 
line is formed over a larger vertical distance, and/or the sound speed
is lower, in the atmosphere of \hr.

Again, surface effects could explain the bisector variation in \acir\ 
and \hr\ if they were pulsating in modes with $\ell \ge 3$ (see
\citenb{Hat96}).  Therefore, we are left with two relatively
straightforward explanations for the amplitude and phase variations:
(i)~the stars pulsate in dipole modes and the variations are caused by
depth effects, or (ii)~they pulsate in $\ell = 3$ modes and the
variations are caused by surface effects.  A more complicated
explanation could involve a combination of standing waves, running
waves, surface effects, distorted modes and blending.

\subsection{Rotational modulation in \hr}

There are many examples of rotational modulation shown in this paper.
The large variation in the shape of the modulation between different
observables was an unexpected result, for which we see three possible
explanations:
\begin{enumerate}
\item The oblique pulsator model is wrong and the observed frequency
  triplet is caused by three different modes.
\item The oblique pulsator model is correct and the variations
  are caused by systematic errors in the measurements.
\item The oblique pulsator model is correct but needs to be
  modified to include effects due to spots or some other mechanism.
\end{enumerate}

In the first case, the relative amplitude between different modes
would be expected to vary for different observables.  This is the
natural explanation for the differences between $\nu_{-1}$ and
$\nu_{+1}$ of the metal-line velocity amplitudes and phases
(Figs.~\ref{fig:ampl-ph-1} and~\ref{fig:ampl-ph+1}) and of the \halp
-width amplitudes and phases (Fig.~\ref{fig:hr3831-bis-wid}).
However, it would require some rotational phase-locking process to
account for the frequency separations being equal to the rotation
frequency.

In the second case, no simple systematic error could explain the
variation in the shape of the rotational modulation.  Any such error
would need to be varying with the rotation.  For instance, a variation
in blending could possibly account for the rotational modulation of
the metal-line amplitudes and phases, but it could not account for the
\halp-width variations.

For the third case, we consider the \halp\ line because it is the
least affected by blending.  First, note that the intensity
modulation, using a filter with a FWHM of about 6\AA, is fitted well
by the rotational modulation derived from the photometric triplet.
This means that the {\em total intensity\/} variations below a height of
about 0.62 are in agreement with the oblique pulsator model.  The
bisector velocities are also in agreement, so it is only the width
variations below 0.62 that are not in agreement with the model.
Perhaps this could be explained by hot spots in the upper atmosphere
that cause significant amplitude and phase differences between
pulsational temperature changes in and around the spots.  As the star
rotates, the effect on the \halp\ profile would vary, leading to a
hybrid model combining spots with an oblique pulsator.

\subsection{Future work}

It is clear that the high-resolution study (R$\ga$40000) of different
metal lines will be necessary to improve the understanding of the
pulsation and the structure of the atmosphere in roAp stars, because
of the problems associated with blending at lower resolution
($R\sim5000$).  It will then be possible to compare the velocity
amplitudes and phases between different atoms and ionization states.
Theoretical work needs to be done to calculate formation depths for
individual lines, including the effects of diffusion and the magnetic
field.  The aim is to build up a coherent picture of the pulsation in
the atmosphere.

Of the results on \acir\ and \hr, the \halp\ profile measurements are
least affected by blending and cannot easily be improved using
high-resolution spectroscopy.  The profile measurements require a good
continuum fit across the \halp\ line, which is about 100\AA\ wide in
these A-stars.  For this reason, a spectrum which is stable across at
least 400\AA\ is preferable.  This is harder to obtain with
high-resolution spectroscopy due to the blaze pattern and the size of
CCD detectors.  Although improvement could be made on the results
using intermediate-resolution spectroscopy ($R\sim$15000), modelling
the formation of the \halp\ line is more important.  Determining depth
and surface effects on the \halp\ profile will complement
high-resolution spectroscopic and theoretical studies of metal lines.

Photometric oscillation spectra could be considerably improved from
observations of roAp stars using small space telescopes.  Two such missions
plan to include roAp stars as part of a project to measure oscillations in
nearby stars ({\em MONS}\footnote{Measuring Oscillations in Nearby Stars\\
({\tt http://www.obs.aau.dk/MONS/})}, see \citenb{Bal98}; and {\em
MOST}\footnote{Microvariability \& Oscillations of STars\\ ({\tt
http://www.astro.ubc.ca/MOST/})}).  If successful, it will result in the
detection of new modes and improved asteroseismology of the brightest roAp
stars.

The spectroscopic study of the pulsations in roAp stars has produced
more questions than answers, due to the complexity of the observed
spectral changes.  For the same reason, the `amount of information'
obtainable is very large (e.g., variations in amplitudes and phases)
compared to other pulsating stars.  Therefore, roAp stars are prime
stellar objects for testing diffusion, magnetic field and pulsation
theories.

\section*{Acknowledgements}

We would like to thank: 
Don Kurtz, Jaymie Matthews and Lawrence Cram for helpful discussions; 
the referee for helpful comments; 
the MSO staff for observing support; 
the SAAO staff for photometric data, and;
the Australian Research Council for financial support.

\bsp
\label{lastpage}

\end{document}